\documentclass[runningheads]{llncs}
\usepackage[utf8]{inputenc}
\usepackage[T1]{fontenc}
\usepackage{graphicx,verbatim}
\graphicspath{ {Image} }
\usepackage{subfigure}
\usepackage{amsfonts}
\usepackage{svg}
\usepackage{amsmath}
\usepackage{booktabs}
\usepackage{tabularx}
\usepackage{gensymb}
\usepackage{makecell}
\usepackage{caption}
\usepackage{subfigure}
\usepackage{float}
\usepackage{marvosym}  
\usepackage{hyperref}
\usepackage{ulem}

\usepackage{colortbl}
\definecolor{lightgreen}{RGB}{144, 238, 144} 
\definecolor{green}{RGB}{189, 230, 205}
\definecolor{myblue}{RGB}{228,238,188}          
\definecolor{gray}{RGB}{255,248,197} 

\usepackage{color}

\begin{document}
\title{ NeRF-based CBCT Reconstruction needs Normalization and Initialization}

\author{
Zhuowei Xu\textsuperscript{‡}\inst{1,2} \and 
Han Li\textsuperscript{‡}\inst{2,3,7} \and 
Dai Sun\inst{1,2} \and
Zhicheng Li\inst{1,2} \and
Yujia Li\inst{1,2,4} \and
Qingpeng Kong\inst{1,2} \and
Zhiwei Cheng\inst{1,2} \and
Nassir Navab\inst{3} \and 
S. Kevin Zhou\inst{1,2,4,5,6}\thanks{Corresponding author: skevinzhou@ustc.edu.cn}}

\institute{
School of Biomedical Engineering, Division of Life Sciences and Medicine, University of Science and Technology of China (USTC), Hefei, 230026, China \and
Center for Medical Imaging, Robotics, Analytic Computing \& Learning (MIRACLE), Suzhou Institute for Advance Research, USTC, Suzhou, 215123, China \and
Computer Aided Medical Procedures (CAMP), TU Munich, 80333, Germany 
 \and
Key Laboratory of Intelligent Information Processing of Chinese Academy of Sciences (CAS), Institute of Computing Technology, CAS, Beijing, 100190, China
\and
Jiangsu Provincial Key Laboratory of Multimodal Digital Twin Technology, Suzhou, 215123, China 
\and
 State Key Laboratory of Precision \& Intelligent Chemistry, USTC, Hefei, China
 \and
Munich Center for Machine Learning (MCML), Munich, Germany
}

\authorrunning{Xu et al.}

\maketitle

\footnotetext[1]{\textsuperscript{‡}These authors contributed equally to this work.}              
\begin{abstract}
Cone Beam Computed Tomography (CBCT) is widely used in medical imaging. However, the limited number and intensity of X-ray projections make reconstruction an ill-posed problem with severe artifacts. NeRF-based methods have achieved great success in this task. However, they suffer from a \textbf{local-global training mismatch} between their two key components: the hash encoder and the neural network. Specifically, in each training step, only a subset of the hash encoder’s parameters is used (\textbf{local sparse}), whereas all parameters in the neural network participate (\textbf{global dense}). Consequently, hash features generated in each step are highly misaligned, as they come from different subsets of the hash encoder. These misalignments from different training steps are then fed into the neural network, causing repeated inconsistent global updates in training, which leads to unstable training, slower convergence, and degraded reconstruction quality.
Aiming to alleviate the impact of this local-global optimization mismatch, we introduce a \textbf{Normalized Hash Encoder}, which enhances feature consistency and mitigates the mismatch. Additionally, we propose a \textbf{Mapping Consistency Initialization(MCI)} strategy that initializes the neural network before training by leveraging the global mapping property from a well-trained model. The initialized neural network exhibits improved stability during early training, enabling faster convergence and enhanced reconstruction performance.
Our method is simple yet effective, requiring \textbf{ only a few lines of code} while substantially improving training efficiency on 128 CT cases collected from 4 different datasets, covering 7 distinct anatomical regions.
The implementation is publicly available at GitHub:\url{https://github.com/iddifficult/NI\_NeRF}.


\end{abstract}
\begin{figure}[ht]
    \centering
    \includegraphics[width=0.9\textwidth]{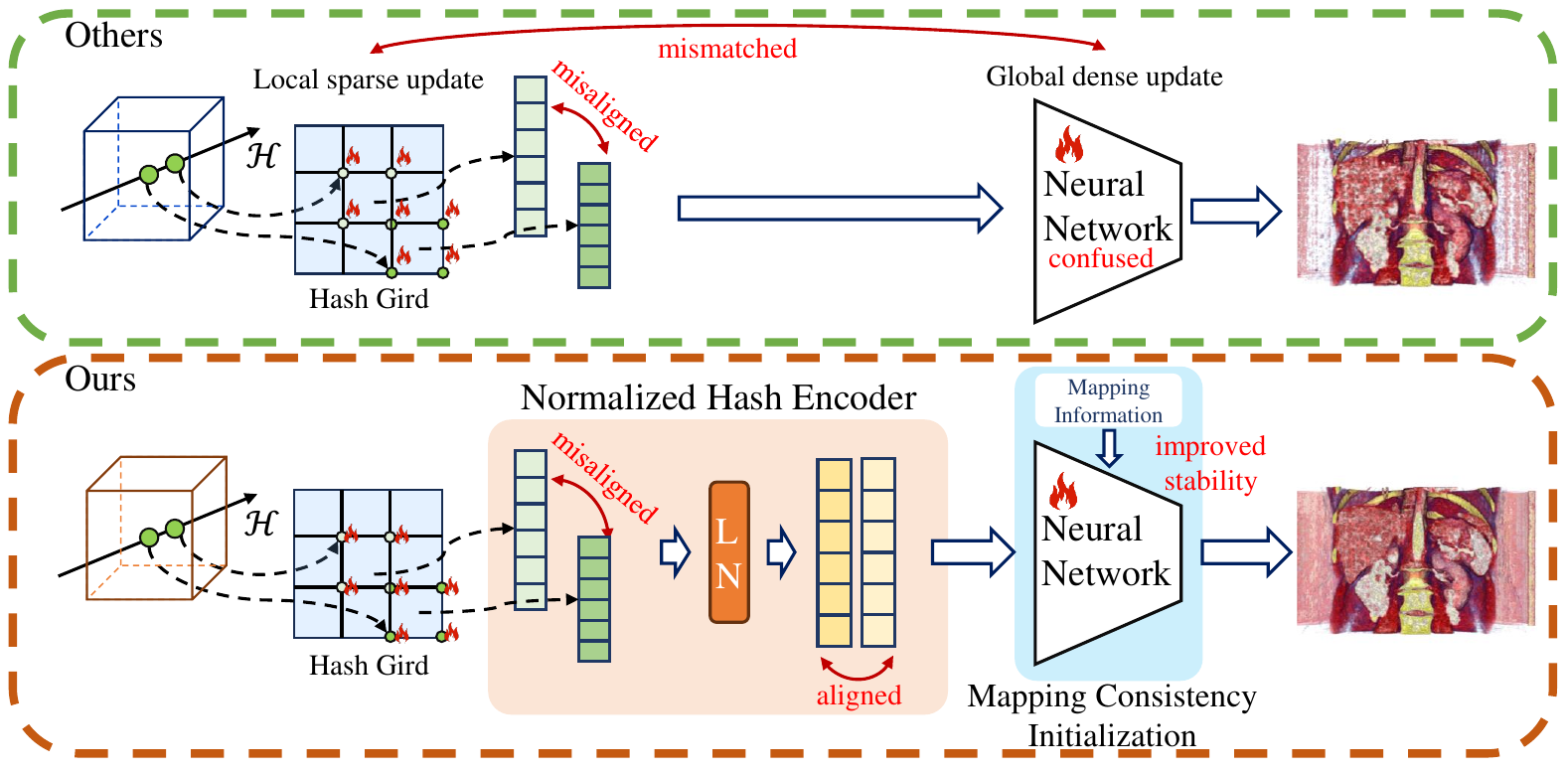}
    \caption{This figure illustrates the update mismatch and feature misalignment problem. The small flame icons indicate the updated sections.} 
    \label{fig:first_result} 
\end{figure}

\section{Introduction}
Cone Beam Computed Tomography (CBCT) is widely used in dental, orthopedic, and interventional imaging due to its lower radiation dose, and faster scanning speed~\cite{scarfe2006clinical}. However, the potentially harmful effects of X-ray radiation limit the intensity and number of projections in CBCT scans, leading to sparse-view data acquisition. This sparsity in projections makes CBCT reconstruction an ill-posed problem, causing image degradation and severe artifacts.

To relieve this issue, numerous sparse-view CBCT reconstruction algorithms have been proposed, broadly categorized into three types: traditional methods such as FDK~\cite{feldkamp1984practical}  and SART~\cite{andersen1984simultaneous}, supervised learning methods like iBP-Net and DIF-Net~\cite{jiao2021dual,lin2023learning,zheng2023ultrasparse,liu2024volumenerf}, and self-supervised methods like NeRF-based ~\cite{zang2021intratomo,ruckert2022neat,cai2024structure,zha2022naf} and 3DGS-based approaches~\cite{zha2024r,cai2024radiative,gao2024ddgs}. Among them, NeRF has emerged as a powerful self-supervised method, effectively reducing artifacts in traditional methods, eliminating the need for paired data in supervised learning, and requiring no distribution assumptions as in 3DGS-based methods that may inherently introduce false artifacts. However, the vanilla NeRF~\cite{mildenhall2020nerf} architecture needs an extremely long training convergence time (hours to days) because it relies entirely on a neural network (i.e., multilayer perceptron or MLP) to learn the mapping of the entire 3D space, resulting in a massive neural network and slow convergence speed.

To address this, recent NeRF-based methods have replaced frequency encoding with hash encoding as the positional encoding, significantly reducing computational complexity and training difficulty, thereby leading to much faster convergence~\cite{zha2022naf,cai2024structure}. While hash encoding effectively accelerates training, it introduces a new challenge: a fundamental \textbf{local-global optimization mismatch} between the hash encoder and the neural network. Existing methods primarily focus on designing efficient models or effective ray sampling strategies but have overlooked this critical issue. Specifically, as shown in Fig.~\ref{fig:first_result}, since the parameters on the hash grid are independently learnable~\cite{mueller2022instant}, only a subset of the hash encoder’s parameters is used (\textbf{local sparse})  in each training step, whereas all parameters in the neural network participate (\textbf{global dense}). Consequently, hash features generated in each step are highly misaligned, as they come from different subsets of the hash encoder. These misaligned features from different training steps are then fed into the neural network, causing repeated inconsistent global neural network updates in each training step, which leads to unstable training, slower convergence, and degraded reconstruction quality.

In this paper, we claim that Normalization and Initialization can alleviate the impact of the local-global optimization mismatch. Therefore, we introduce a \textbf{Normalized Hash Encoder}, which enhances feature consistency and mitigates the mismatch. Specifically, we add a Layer Normalization (LN) between the hash encoding model and the neural network. This ensures that the features of the hash encoding maintain a unified global mean and variance across the whole training process, thereby mitigating the misalignment problem.  Additionally, we propose a \textbf{Mapping Consistency Initialization(MCI)} strategy that initializes the neural network before training by leveraging the global mapping property from a well-trained model. Sepcifically, we first train a complete NeRF-based CBCT reconstruction model on entire volume of one case and then reuse its neural network component as the initialization for other reconstruction tasks. By transferring learned knowledge across cases, the initialized neural network exhibits improved stability during early training, significantly accelerates convergence and enhances reconstruction performance.

To the best of our knowledge, \underline{we are the first to systematically} investigate the \textbf{local-global optimization mismatch} and propose a simple-yet-effective and feasible method. We conduct extensive experiments on 128 CT cases collected from 4 different datasets, covering 7 distinct anatomical regions. The results show that our method not only outperforms NeRF-based methods in terms of reconstruction speed and quality but also achieves comparable reconstruction speeds to 3D Gaussian Splatting (3DGS) while surpassing 3DGS in reconstruction quality.

\section{Method}

\begin{figure}[ht]
    \centering
     \includegraphics[width=\textwidth]{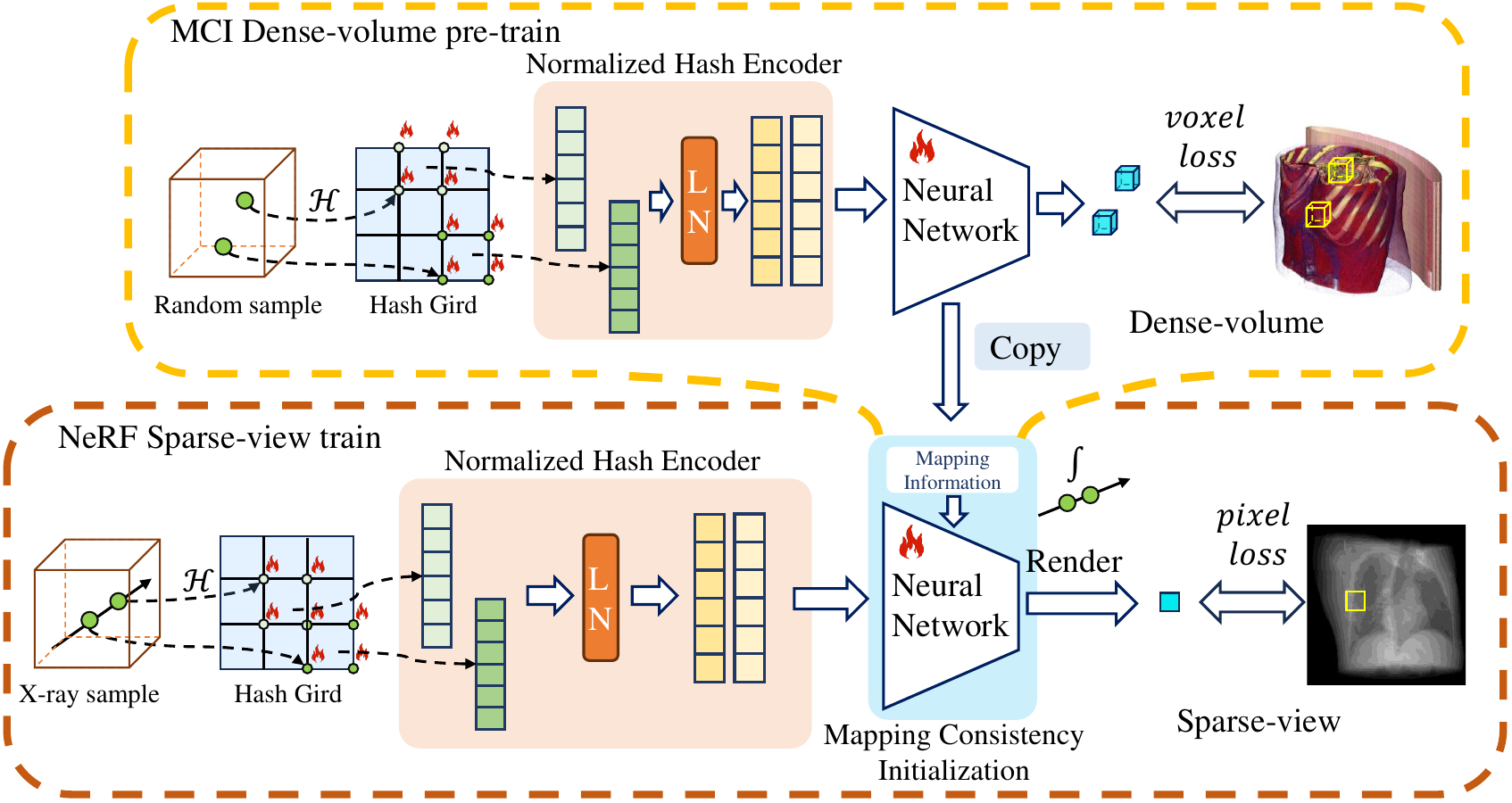}
    \caption{Our pipeline of pre-training and normal reconstruction.} 
    \label{fig:pipline} 
\end{figure}

\subsection{Pipeline}
As shown in Fig.~\ref{fig:pipline}, we display the complete pre-training method and the training process during reconstruction. During pre-training, since the use of ground truth information is permitted, We perform dense random sampling on the entire volume and directly supervise the entire NeRF model using the values corresponding to the sampled points on the GT. During reconstruction, we only load the Layer Normalization (LN) and the neural network weights into the new NeRF model, sample spatial points along the propagation path of the X-ray, and compute the loss by integrating and comparing the values with those corresponding to the projection image.

\subsection{X-ray rendering process}
Due to the penetrative nature of X-rays and the known positions of the light source and detector, we adopt the Beer-Lambert Law (BLL) to replace the \(\alpha\)-blender
 in the original NeRF. The BLL describes the exponential attenuation of light intensity as X-rays pass through an object.Where \(A\) is the projection value, \(I_0\) is the initial X-ray intensity, \(\mu_i\) is the attenuation coefficient at point \(i\), and \(\delta_i\) is the step size.
\begin{equation}
A \doteq -\ln\left(\frac{I}{I_0}\right) = \sum_{i=1}^{N} \mu_i \delta_i.
\end{equation}

\subsection{Normalized Hash Encoder}

\begin{figure}[t]
\centering
\begin{minipage}[t]{0.43\linewidth}
\centering
\includegraphics[width=\linewidth]{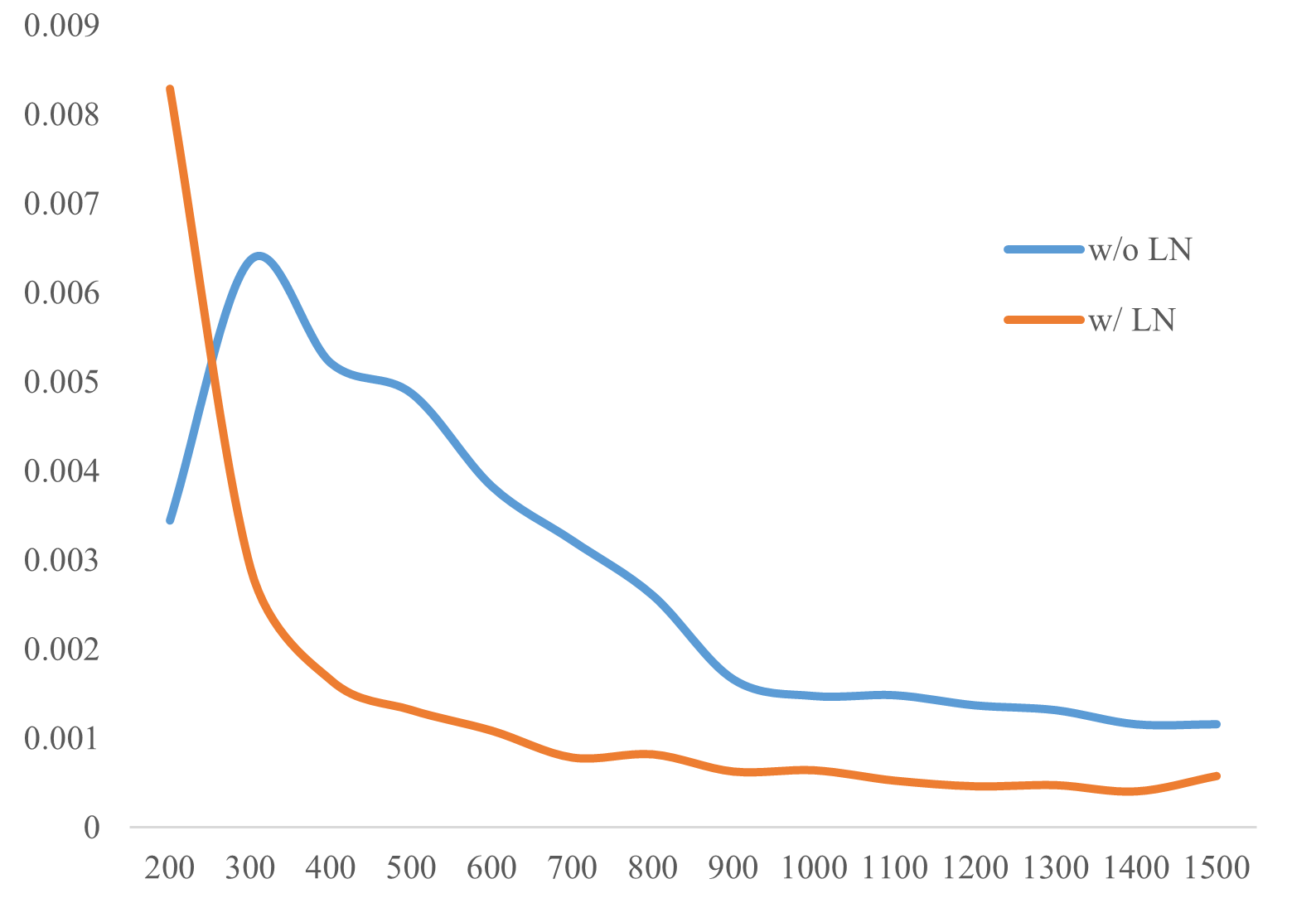}
\caption{Variation of MLP during training process, lower means stabler.}
 \label{fig:variation}
\end{minipage}
\hspace{4mm}
\begin{minipage}[t]{0.43\linewidth}
\centering
\includegraphics[width=\linewidth]{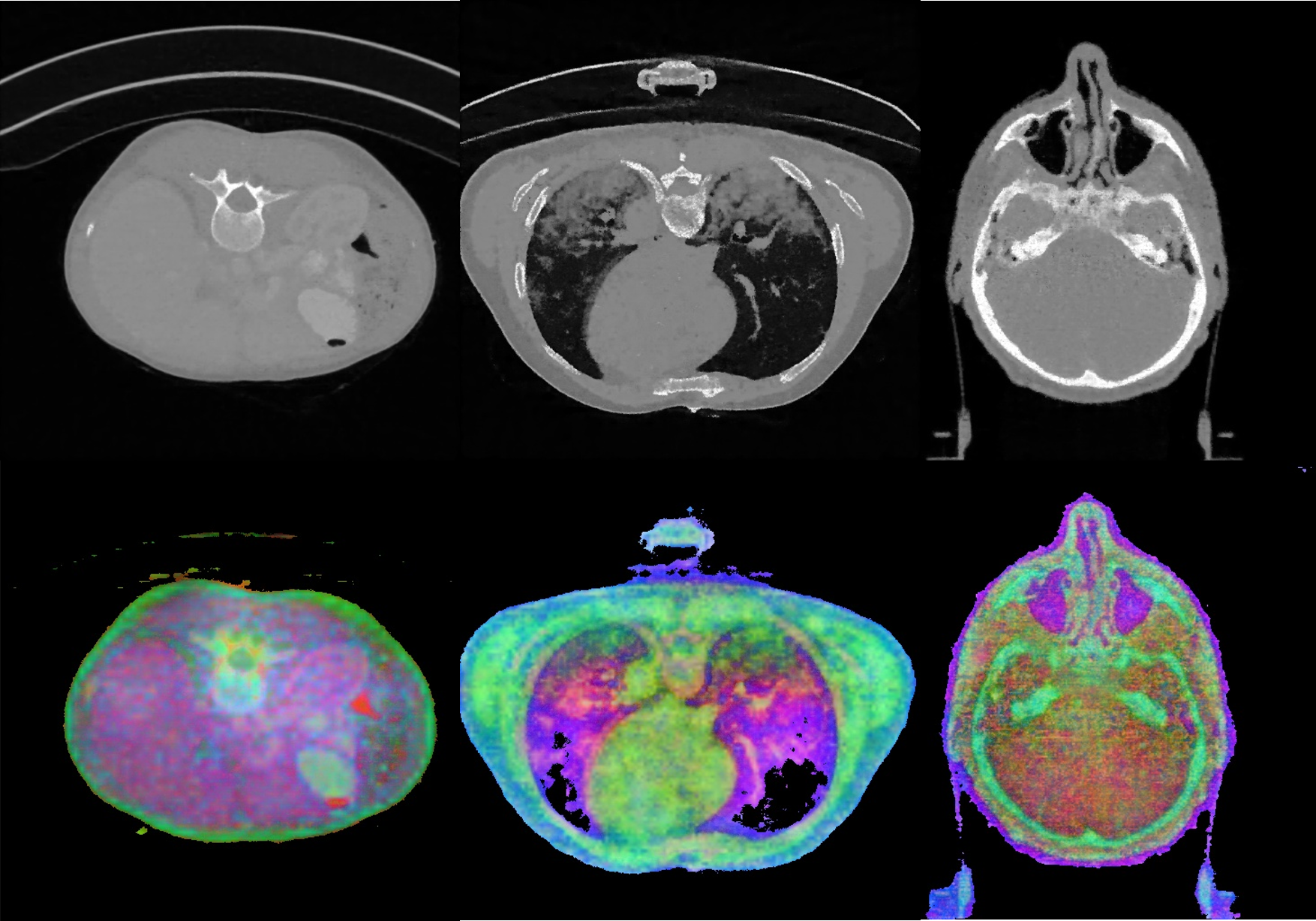}
\caption{Feature PCA results of abdomen, chest and head.}
 \label{fig:PCA}
\end{minipage}
\vspace{-1em}
\end{figure}

Layer Normalization (LN) scales all feature vectors within a batch to have a unified mean and variance. The formula for this transformation is as follows. To validate the impact of feature misalignment on the neural network, we record a batch of hash-encoded features every 100 epochs, along with the network's corresponding outputs. After 100 additional epochs, we reprocess the recorded features and compute the L1 error between the two inference results to assess neural network stability. As shown in Fig.~\ref{fig:variation}, without Layer Normalization (LN), the neural network converges slowly and exhibits significant confusion; incorporating LN enables the network to converge quickly and stably. We place LN between the neural network and the hash encoder.

Furthermore, at the channel level of the hash features, we find some channels exhibit meaningless noise. Via Fast Fourier Transform (FFT), we distinguish and mask these channels during training to improve model's performance.

\subsection{Mapping consistency initialization}

To further investigate the function of neural network in NeRF, we employ Principal Component Analysis (PCA) to analyze the features of the hash encoding. Specifically, We use the first three principal components as the RGB channels of the image, where similar colors indicate similar features. As shown in Fig.~\ref{fig:PCA} spatial points with similar attenuation coefficients(exhibit similar brightness in CT) have similar feature vectors. This finding indicates that the neural network inherently learns a consistent simple mapping in different cases, which inspired us to employ pre-training to facilitate faster and more stable model training.

To avoid the significant time consumed by traditional NeRF training, we bypass the NeRF rendering pipeline and directly supervise the attenuation coefficients of individual spatial points using a voxel-to-voxel loss, enabling an efficient dense-volume pre-training process. \(\mathcal{M}\) means NeRF, x,y,z means coordinates.
\begin{equation}
Voxel\_loss = \| \mu_{gt}-\mathcal{M}(x,y,z)\|_1.
\end{equation}

\subsection{Sparse-view training}
During training, we sample spatial points along X-ray trajectories and compute predicted projection values using the BLL. Then minimize the L1 loss between predicted and GT projections.
\begin{equation}
Pixel\_loss = \| A_{gt}-\sum_{i=1}^{N}\mathcal{M}(x_i,y_i,z_i) \delta_i\|_1.
\end{equation}

\section{Experiments}
\subsection{Settings}
\textbf{Data} We evaluate our method on four public datasets, covering common medical CT scenarios, including the head, chest, and abdomen. For the chest, we use the Covid-19 dataset~\cite{antonelli2022medical}, which includes data from 10 patients. For the abdomen, we utilize the Pancreas\_CT dataset~\cite{roth2016data}, comprising 82 patients. For the head, we extract 34 cases from the HAN\_seg~\cite{podobnik2023han} dataset. All these datasets have a resolution of $512 \times 512 \times \text{num\_slice}$.  Furthermore, we test four medical case in R\textsuperscript{2}\_Gaussian dataset, which is in resolution of $256 \times 256 \times \text{num\_slice}$. To validate our method, we use 50 views for each dataset and conduct our projections with TIGRE toolbox~\cite{biguri2016tigre} following NAF setting~\cite{zha2022naf}.

\noindent\textbf{Baselines}
We compared our method with five baselines: FDK~\cite{feldkamp1984practical}, a traditional analytical method; SART~\cite{andersen1984simultaneous}, a widely used iterative algorithm for sparse-view CT reconstruction; NAF~\cite{zha2022naf}, a NeRF variant with hash encoding; SAX-NeRF~\cite{cai2024structure}, a high-quality but time-consuming CT reconstruction method; and R\textsuperscript{2}\_Gaussian~\cite{zha2024r}, a 3DGS-based approach.

\begin{figure}[t]
    \centering
     \includegraphics[width=\textwidth]{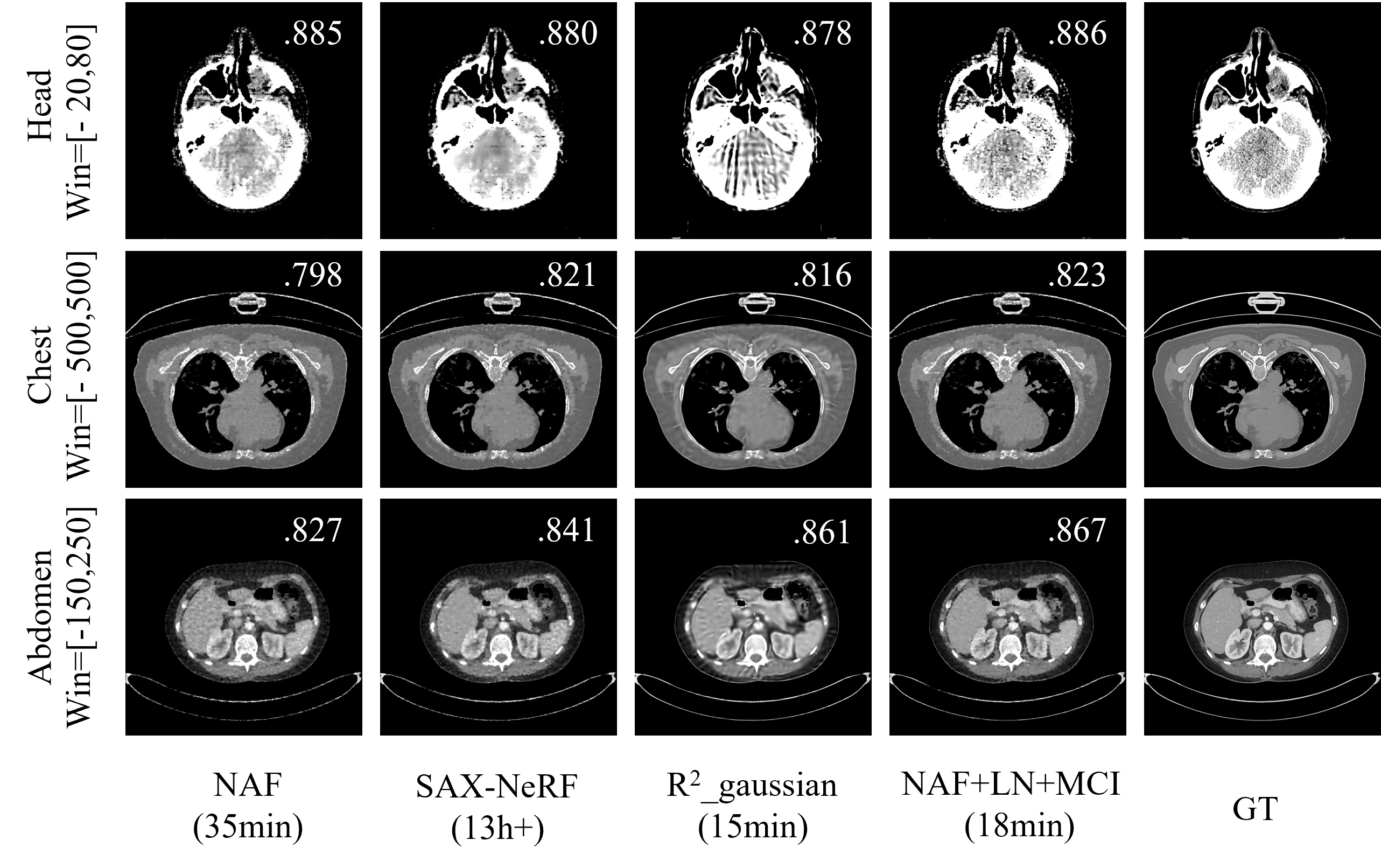}
    \caption{Visualization of Reconstruction Results. The SSIM is displayed in the upper right corner of each image.} 
    \label{fig:slice_all} 
\end{figure}
\noindent\textbf{Implementation details}
Our experiments are implemented in PyTorch~\cite{NEURIPS2019_9015} and CUDA~\cite{cuda}, trained using the Adam optimizer~\cite{diederik2014adam} with a learning rate of $1 \times 10^{-3}$. The batch size of view is 1, and the number of sample rays per projection is 1024, with 320 sample points per ray. We only use one abdomen case to pre-train neural network and load the same weights in all cases. For each method involved in the testing, we train it for a sufficiently long period (3000 epochs for the NeRF-based method and 30000 epochs for R\textsuperscript{2}\_Gaussian) to ensure convergence.

\subsection{Reconstruction performance}
\begin{figure}[ht!]
\centering
\begin{minipage}[t]{0.47\linewidth}
\centering
\includegraphics[width=\linewidth]{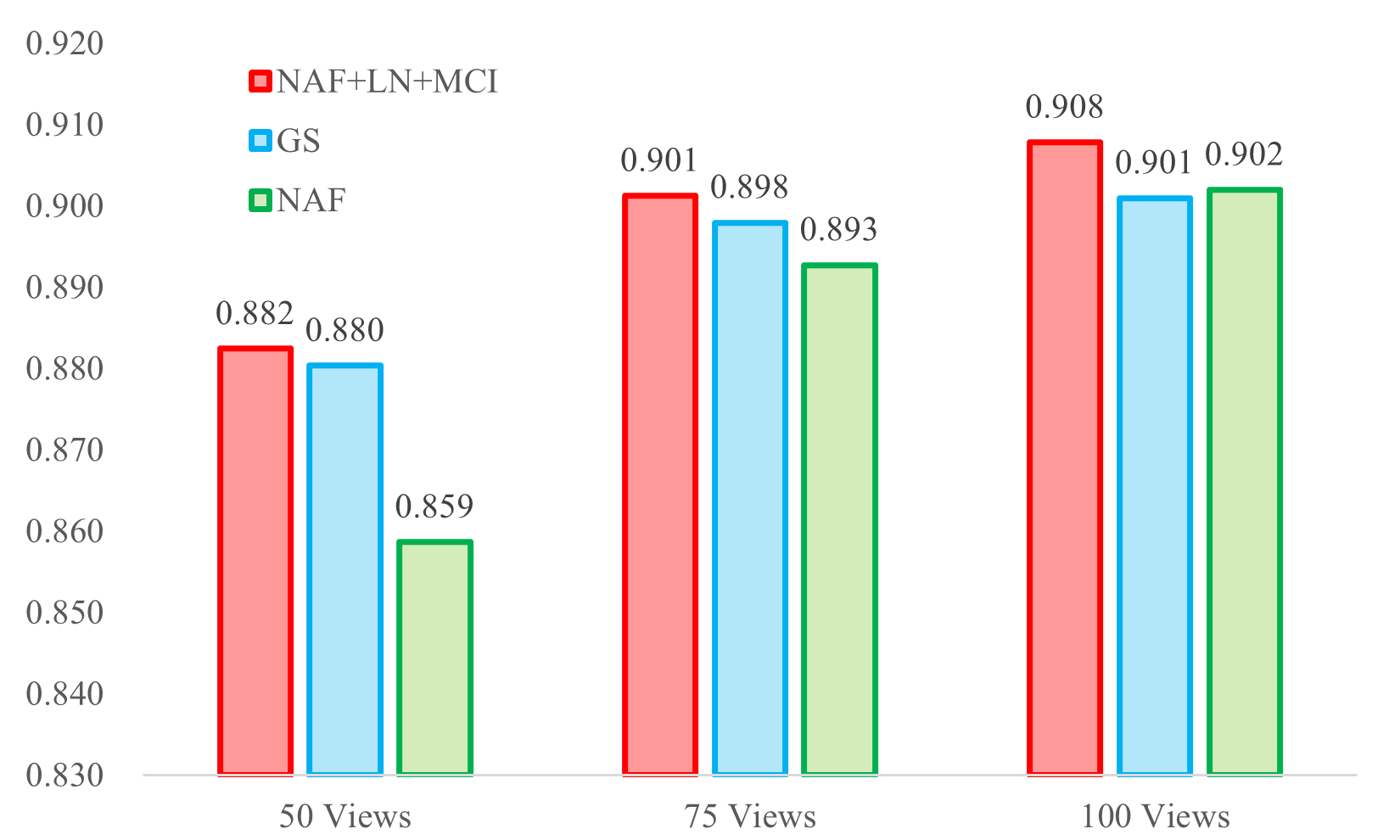}
\caption{SSIM score of different methods using different number of views.}
 \label{fig:multi-view}
\end{minipage}
\hspace{4mm}
\begin{minipage}[t]{0.47\linewidth}
\centering
\includegraphics[width=\linewidth]{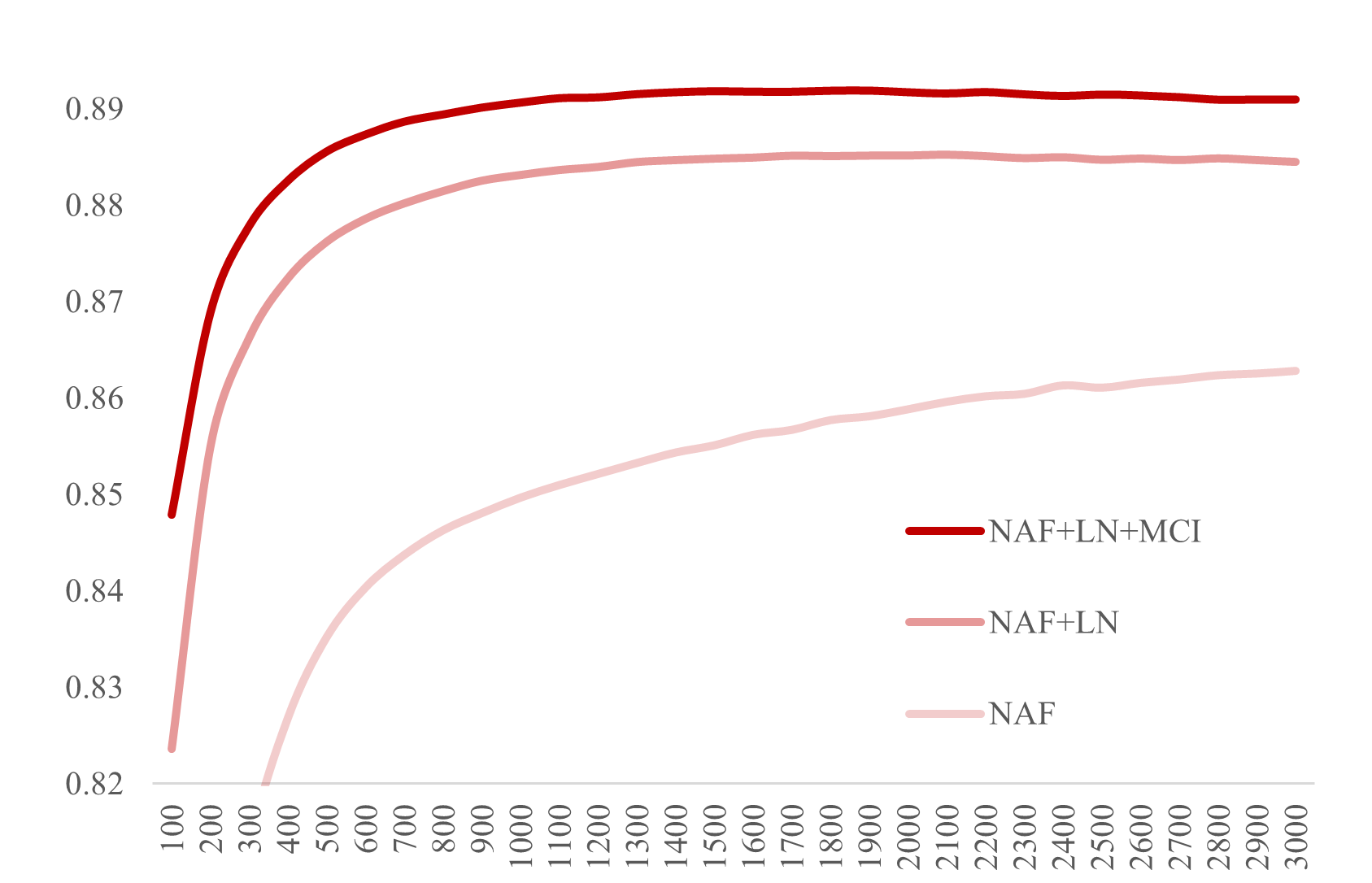}
\caption{Ablation study of LN and MCI on NAF.}
 \label{fig:naf ablation}
\end{minipage}
\vspace{-1em}
\end{figure}

Our evaluation includes two parts: (1) traditional image quality metrics (PSNR and SSIM~\cite{wang2004image}), and (2) Average Segment Dice Score. For the latter, we use TotalSegmentator~\cite{wasserthal2023totalsegmentator,isensee2021nnu}, a widely adopted segmentation model, to assess the similarity between ground truth and reconstructed images~\cite{guo2024maisi}. Additionally, we test our method's performance under different numbers of views on three cases\ref{fig:multi-view}.
\setlength{\fboxsep}{0pt}
\begin{table}[t]
\begin{minipage}{\linewidth}
\centering
\caption{PSNR/SSIM score of methods on 4 datasets. \colorbox{green}{Best} and \colorbox{myblue}{second-best}.}
\label{tab:big_publish_dataset}
\begin{tabular}{m{0.25\columnwidth}<{\centering}|m{0.23\columnwidth}<{\centering}m{0.23\columnwidth}<{\centering}m{0.23\columnwidth}<{\centering}}
\toprule
Method & Chest~\cite{antonelli2022medical} & Abdomen~\cite{roth2016data} & Head~\cite{podobnik2023han} \\
\midrule
FDK~\cite{feldkamp1984practical} & 19.56/.2867 & 21.78/.3865 & 24.62/.3067 \\
SART~\cite{andersen1984simultaneous} & 23.75/.5855 & 28.05/.7645 & 30.42/.8821 \\
R\textsuperscript{2}\_Gaussian~\cite{zha2024r} & \colorbox{myblue}{27.40}/\colorbox{myblue}{.7547} & 34.30/\colorbox{myblue}{.9192} & \colorbox{green}{35.38}/\colorbox{green}{.9637} \\
NAF~\cite{zha2022naf} & 26.12/.7149 & 33.53/.8961 & 34.09/.9533 \\
\cline{2-4}

NAF+LN+MCI & \colorbox{green}{27.51}/\colorbox{green}{.7618} & \colorbox{green}{35.54}/\colorbox{green}{.9234} & \colorbox{myblue}{34.53}/\colorbox{myblue}{.9579} \\
\bottomrule
\end{tabular}
\label{tab:GS_medical_dataset}
\begin{tabular}{m{0.25\columnwidth}<{\centering}|m{0.1725\columnwidth}<{\centering}m{0.1725\columnwidth}<{\centering}m{0.1725\columnwidth}<{\centering}m{0.1725\columnwidth}<{\centering}}
Method & GS\_chest & GS\_foot & GS\_head & GS\_jaw \\
\midrule
FDK~\cite{feldkamp1984practical} & 26.28/.4967 & 26.22/.4479 & 29.35/.5753 & 29.73/.6524 \\
SART~\cite{andersen1984simultaneous} & 31.87/.8652 & 30.29/.8669 & 35.18/.9252 & 33.13/.8388 \\
R\textsuperscript{2}\_Gaussian~\cite{zha2024r} & \colorbox{myblue}{36.27}/\colorbox{myblue}{.9482} & \colorbox{green}{31.98}/.8813 & 41.26/.9842 & \colorbox{green}{36.40}/\colorbox{green}{.8885} \\
SAX\_NeRF~\cite{cai2024structure} & 35.88/.9347 & \colorbox{myblue}{31.97}/.8828 & 41.11/.9814 & \colorbox{myblue}{35.37}/.8707 \\
NAF~\cite{zha2022naf} & 34.77/.9050 & 31.3/.8726 & 40.65/.9736 & 34.15/.8366 \\
\cline{2-5}

NAF+LN+MCI & \colorbox{green}{37.18}/\colorbox{green}{.9484} & 31.64/\colorbox{myblue}{.8861} & \colorbox{green}{42.85}/\colorbox{green}{.9885} & 35.36/\colorbox{myblue}{.8745} \\
\bottomrule
\end{tabular}
\end{minipage}
\end{table}
\setlength{\fboxsep}{3pt}

\begin{figure}[ht!]
    \centering
    \includegraphics[width=\textwidth, keepaspectratio]{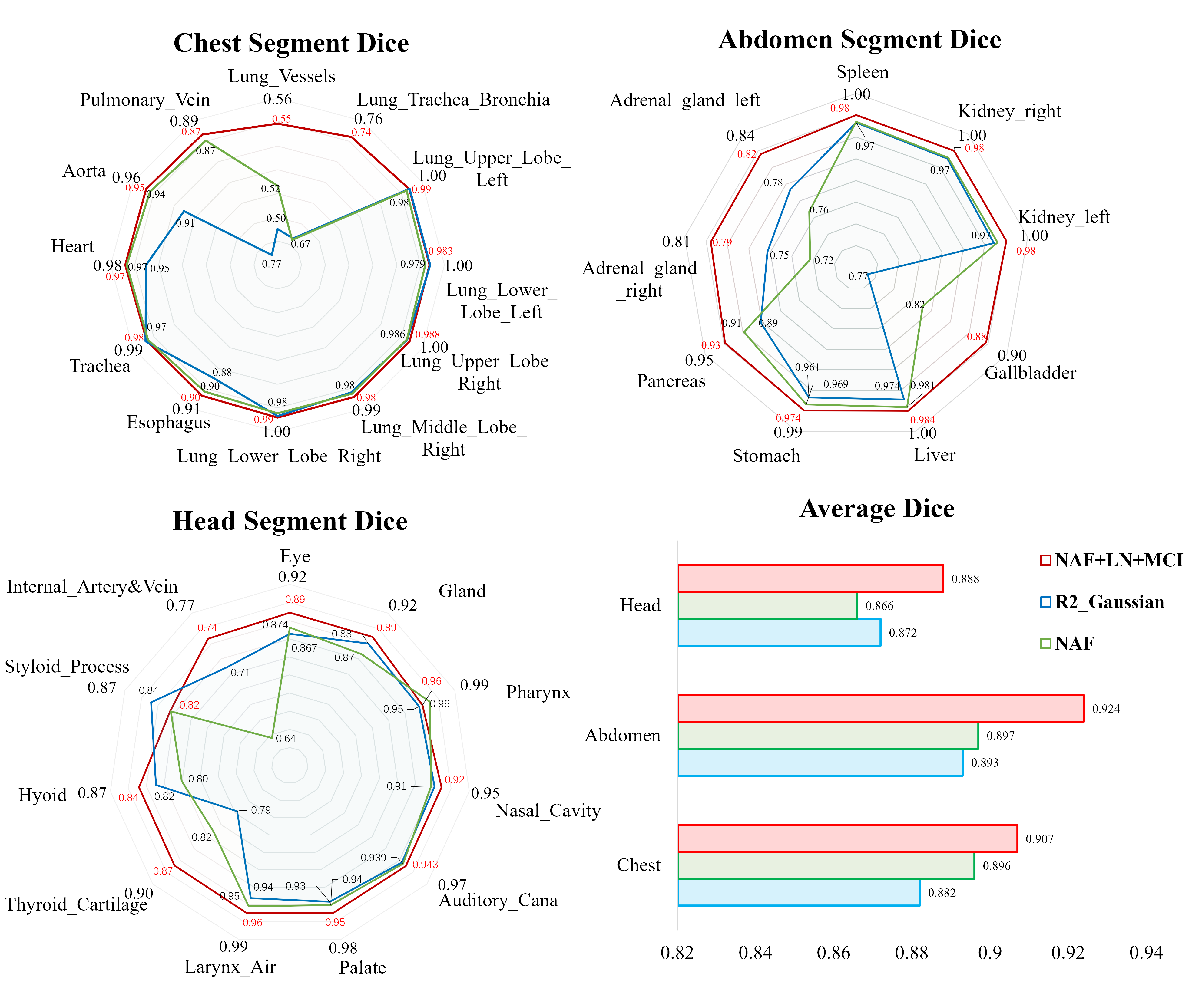}
    \caption{The segment results of different methods on 3 datasets.} 
    \label{fig:total_seg} 
\end{figure}

\noindent\textbf{Image quality performance} 
Through our method, NAF achieves better image quality and the fastest reconstruction speed among NeRF-based approaches, as in Table~\ref{tab:big_publish_dataset}. Compared to Gaussian Splatting (GS)-based methods, our method shows superior quality on most datasets and has a similar converge time (18 min). To further validate the clinical relevance of these methods, we visualize the images using different window settings, closely mimicking real clinical scenarios. The results show that GS-based methods exhibit severe artifacts and blurred organ boundaries as in Fig.~\ref{fig:slice_all}.

\noindent\textbf{Segment quality performance} 
Our method makes NAF outperform much better than R\textsuperscript{2}-Gaussian on segment Dice, as in Fig.~\ref{fig:total_seg}. These results demonstrate that our method helps NAF's reconstruction results have better anatomical structure fidelity.

\subsection{Ablation study}
To evaluate the impact of our proposed improvements, we select three images and compute the average SSIM across them. As shown in Fig.~\ref{fig:naf ablation}, each modification not only accelerates convergence but also enhances the final performance.

\section{Conclusion}
In conclusion, we address the local-global training mismatch in NeRF-based CBCT reconstruction, where the hash encoder's local sparse updates conflict with the neural network's global dense updates, causing misaligned features, unstable training, and slow convergence. To resolve this, we propose a Normalized Hash Encoder for feature consistency and a Mapping Consistency Initialization(MCI) strategy for stable neural network initialization. Our method, requiring minimal code changes, but significantly improves training efficiency and reconstruction quality. This work provides a robust solution for efficient and accurate CBCT reconstruction.

%
%
%
\bibliographystyle{splncs04}
\bibliography{main}

\end{document}